\documentclass[aps,pra,reprint,groupedaddress,showpacs]{revtex4-1}

\usepackage{graphicx}
\usepackage{amsmath}
\usepackage{setspace}
\usepackage{dcolumn}
\usepackage{bm}
\usepackage{color}

\begin{document}

\title{  Nondipole effects in atomic dynamic interference }

\author{Mu-Xue Wang, Hao Liang, Xiang-Ru Xiao, Si-Ge Chen}
\affiliation{State Key Laboratory for Mesoscopic Physics  and  Collaborative Innovation Center of Quantum Matter, School of Physics,  Peking University, Beijing 100871, China}

\author{Wei-Chao Jiang}
 \email[]{jiang.wei.chao@szu.edu.cn}
\affiliation{College of Physics and Energy, Shenzhen University, Shenzhen 518060, China}

\author{Liang-You Peng}
 \email[]{liangyou.peng@pku.edu.cn}
\affiliation{State Key Laboratory for Mesoscopic Physics  and  Collaborative Innovation Center of Quantum Matter, School of Physics,  Peking University, Beijing 100871, China}

\affiliation{Collaborative Innovation Center of Extreme Optics, Shanxi University, Taiyuan, Shanxi 030006, China}


\date{\today}

\begin{abstract}

\par Nondipole effects in the atomic dynamic interference are investigated by numerically solving  the time-dependent Schr\"odinger equation~(TDSE) of hydrogen.
It is found that the inclusion of nondipole corrections in the TDSE can induce   momentum shifts of photoelectrons in the opposite direction of the laser propagation.   The magnitude of the momentum shift is roughly proportional to the laser peak intensity and to the  momentum component of the photoelectron along the laser propagation. By including the nondipole corrections of the Volkov phase into a semi-analytical model previously developed under the dipole approximation,  all the main features of the momentum shifts can be nicely reproduced. Through an analytic expression, the origin of such momentum shifts is attributed to the nondipole phase difference between the two electron wave packets ejected in the rising edge and the falling edge, which will interfere with each other and result in the final fringe pattern.
One important consequence of such momentum shifts is that they can smooth out   the peak splitting induced by the dynamic interference  in the photoelectron energy spectrum. Nevertheless, it should be emphasized that the dynamic interference persists in the photoelectron momentum distributions and is not suppressed at all for the laser parameters considered in this work.

\end{abstract}


\maketitle

\section{INTRODUCTION}

Interference processes are at the heart of the optical and quantum mechanical phenomena, which can be realized through a spatial or temporal double-slit prototype. The availability of different new light sources has enabled the possibility to observe new types of quantum interference in atomic and molecular systems.  For example,   free-electron lasers~(FELs)~\cite{Emma2010, McNeil2010, Ishikawa2012} can now provide  x-rays at wavelengths down to ~1 \AA ngstrom with an unprecedented intensity around $10^{20}$ W/cm$^{2}$)~(see, e.g, a recent review in Ref.~\cite{JPB2018} and reference therein).  For laser pulses at large photon energies and high intensities, the so-called ``dynamic interference" in the atomic ionization has been theoretically observed~\cite{Toyota2008b, Toyota2008a, Demekhin2012, Demekhin2012a, Yu2013, Toyota2015b, Toyota2016, Artemyev2016, Baghery2017a, Ning2018}. It refers to the interference between two electronic wave packets respectively ejected in the rising  and the falling edge of a laser pulse.

According to the Einstein's photoelectric law,  a single peak at $E=\omega-I_p$  is expected in the photoelectron energy spectrum for   one-photon ionization at a low laser intensity.  However,  the single peak can gradually evolve into a multi-peak structure due to the dynamic interference when the laser intensity is increased.
The conditions to observe such a dynamic interference are   recently discussed~\cite{Baghery2017a,Jiang2018}. In fact, to observe the dynamic interference  in the ground state of hydrogen at a moderately high laser frequency~($\sim$50 eV),
the peak intensity needs to be high enough (above $10^{18}$ W/cm$^2$) for the atomic stabilization to occur~\cite{Jiang2018}.
The relative theoretical  studies are presently limited to the dipole approximation for the laser-atom  interaction.
However, one expects that the nondipole correction may play a non-negligible role  at large photon energies with such a high laser intensity~\cite{Reiss2014b, Forre2014a}.
For instance, the inclusion of the nondipole corrections may weaken the atomic stabilization~\cite{Kylstra2000a, Physics2002a, Popov2003b, Forre2005a, Emelin2014a, Simonsen2015, Staudt2003, Staudt2006a, Emelin2017}.

In the dipole approximation, the time-dependent Hamiltonian is cylindrically symmetric  for a
linearly  polarized laser pulse.
However, the inclusion of the leading-order  corrections of the nondipole effects  will break the cylindrical symmetry of the Hamiltonian. As a result, the cylindrical symmetry of the angular distribution of the photoelectrons can also  be broken when  the nondipole corrections  become non-negligible.
In some cases, the asymmetrical angular distribution of the photoelectrons can be treated as a result of the interference  between
the dipole ionization paths and the nondipole ionization paths.  Actually, there is a long history in calculating and measuring the asymmetrical distribution parameters~\cite{Krassig1995, Martin1998, Dolmatov1999, Derevianko2000, Hemmers2001, Krassig2002, Hemmers2003, Hemmers2006}. Recently, several theoretical studies discussed the nondipole asymmetrical angular distributions~\cite{Forre2006, Forre2006a, Zhou2013a, Forre2007a, Moe2018, Forre2014a}.
A related topic is the photon-momentum transfer and partition in the atomic and molecular ionization~\cite{Smeenk2011b, Titi2012, Klaiber2013, Reiss2013, Liu2013, Yakaboylu2013, Ludwig2014a, Chelkowski2014b, Chelkowski2015a, Ivanov2015, Cricchio2015, Tao2017, He2017, Chelkowski2017b, Wang2017, Lao2016, Chelkowski2018}.
Here, the momentum of the photon can be related to the average value of the photoelectron momentum  along the laser propagation direction, whose  nonzero value   is an appearance of the broken cylindrical symmetry.

In this work, we particularly investigate  the nondipole effects in the  dynamic interference by numerically solving the time-dependent Schr\"odinger equation~(TDSE).
We present detailed angularly distinguished  momentum spectra of the photoelectron within/byeond the dipole application.
In the dipole approximation,  the  ring structures  resulted from the dynamic interference in the momentum spectra are  concentric at a zero momentum and symmetric about the line of the laser polarization according to the dipole transition.
For the nondipole calculations, ring-like structures caused by the dynamics interference are still clearly present,  but their geometric centers are apparently shifted in the opposite direction of the laser propagation and the rings are asymmetric  about the axis of the laser polarization, which is due to the leading-order consideration of the magnetic force of the pulse.     Such a   shift in the momentum spectra can
significantly suppress the peak splitting in the angularly integrated energy spectrum.
We further explore the origin of the momentum shift and find that it is closely related to
the nondipole phase differences between electrons emitted in
the rising and falling edge of the laser pulse.
By including  the  nondipole phase term
into a semi-analytical model previously developed, the momentum shifts of the interference rings  can be  nicely reproduced.

\section{Methods}

In the nonrelativistic situation, the dynamics of a hydrogen atom interacting with a classical electromagnetic field can be  described by the time-dependent Schr\"odinger equation~(atomic units are employed throughout unless otherwise stated),
\begin{equation}\label{Sequa}
  {i}\frac{\partial}{\partial t}\Psi(\mathbf{r},t)={H}\Psi(\mathbf{r},t),
\end{equation}
where the full Hamiltonian in the  velocity gauge is given by
\begin{equation}\label{FullH}
  H=\frac{1}{2}\left[\mathbf{p}+\mathbf{A}(\mathbf{r},t)\right]^2+V(\mathbf{r}).
\end{equation}

 With the consideration of the lowest-order nondipole correction~\cite{Walser2000, Kylstra2001a, Forre2014a, Cricchio2015}, the time- and space-dependent vector potential $\mathbf{A}(\mathbf{r},t)$ for a laser pulse linearly polarized along the $z$ axis with a propagation direction in the positive $x$ axis can be written as
 \begin{eqnarray}
     A_z(x,t)= A_z(t)+\frac{x}{c}E_z(t),  \label{Az}
\end{eqnarray}
  in which $A_z(t)$ is the vector potential usually adopted in the dipole approximation and $E_z(t)$ is the corresponding electric field,   with $c$ being the vacuum light speed. Inserting  Eq.~(\ref{Az})   into Eq.~(\ref{FullH}), one   obtains the leading-order corrected nondipole Hamiltonian,
\begin{eqnarray}\label{nodipoleFullH}
  \nonumber H_{\mathrm{nondipole}} &=& -\frac{1}{2}\nabla^2+V(r)-iA_z(t){\partial}_z\\[3mm]
  \nonumber   & & - i\frac{x}{c}E_z(t){\partial}_z+\frac{x}{c}A_z(t)E_z(t),
\end{eqnarray}
 in which we assume a  hydrogen system, with  a spherically symmetric potential $V(\mathbf{r})=V(r)=-1/r$.

\par To solve the TDSE, we expand the wave function in terms of the spherical harmonics for the angular coordinates. The resultant  Schr\"odinger equation for the radial wave function can be solved using various discretization methods~\cite{Peng2006a, PENG2015,Jiang2017}. In the present work, we use the finite difference for the radial coordinate $r$ and  the split-operator technique  for the short-time propagator~\cite{Bauer2006a}. The initial state is obtained by an imaginary time propagation in the absence of the external field until the ground-state energy is fully converged. And the final wave function ${\Psi}_{{f}}$ after the end of the external fields is projected onto the scattering states ${\Psi}_{\mathbf{p}}^{(-)}$~\cite{Starace1982} to obtain the differential distribution of the photoelectron with a momentum $\mathbf{p}$, i.e.,
\begin{eqnarray}D(\mathbf{p}) = D(p, \theta, \phi) = |\langle {\Psi}_{\mathbf{p}}^{(-)}|{\Psi}_f \rangle|^2, \end{eqnarray}
 where $\theta$ is the angle between the photoelectron emission direction and the $z$-axis.  To obtain the angularly integrated energy spectrum $D(E)$, one  can change from the momentum  to the energy space according to their Jacobian matrix and carry out the following integration,
\begin{equation}\label{Espectrum}
  D(E)=\int\!\!\int D(\mathbf{p}) {|\mathbf{p}|}\sin{\theta}d\theta d\phi.
\end{equation}

\section{Results and Discussions}

In this section, we will present our main results, which are calculated for a linearly polarized Gaussian-shaped pulse with a carrier frequency of $\omega=53.6$~eV and with a full width at half maximum~(FWHM) about 7 cycle. We will first examine the photoelectron energy distribution $D(E)$ at various laser intensities, where the dynamic interference   gradually appears and drastic  differences are observed between the dipole and the nondipole results at higher intensities. One may suspect that the dynamic interference is severely suppressed by the nondipole effects.

However, as one turns to compare the differential momentum distributions at the highest intensity used~($1\times 10^{19}~\mathrm{W/{cm}^2}$), one finds that nice interference  still persists but with momentum shifts of the ring structures in the opposite direction of the laser propagation. Finally, by including the nondipole phase shift into our previous semi-analytical model based on the dipole approximation, we can reproduce the momentum shifts and find that they smooth out the dynamic interference in  the angularly integrated spectrum $D(E)$.

 For all the results presented in the following,    full convergences  are guaranteed  against the change of the temporal and spatial parameters.  Typically,  the angular momentum $l$ is taken up to 40 and the maximum radial coordinate $r_{\mathrm{max}}\approx 2800$ a.u.

\subsection{Nondipole effects in the energy spectrum}

\begin{figure}
  \centering
 \includegraphics[width=0.9\linewidth]{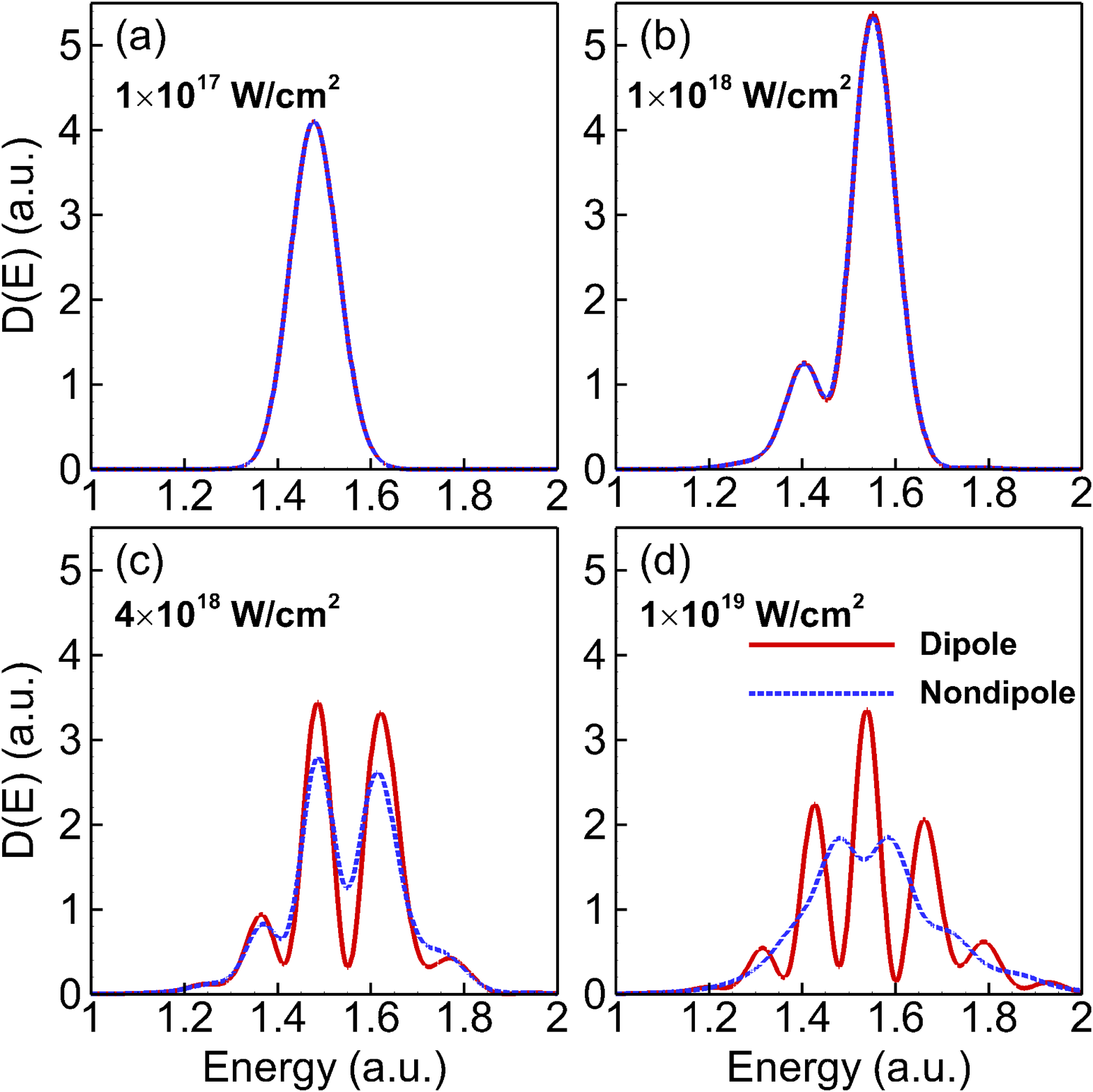}\\
  \caption{(Color online) The photoelectron energy distribution $D(E)$ for the 1s hydrogen state exposed to  linearly polarized Gaussian-shaped pulses with a carrier frequency of $\omega = 53.6$ eV and FWHM of 7 cycle at four different laser intensities: (a) $I_0 = 1.0 \times 10^{17}~\mathrm{W/{cm}^2}$, (b) $I_0 = 1.0 \times 10^{18}~\mathrm{W/{cm}^2}$, (c) $I_0 = 4.0 \times 10^{18}~\mathrm{W/{cm}^2}$, and (d) $I_0 = 1.0 \times 10^{19}~\mathrm{W/{cm}^2}$. Red solid line: the dipole result; blue dotted line: the nondipole result. } \label{fig1}
\end{figure}

Previous studies on  the dynamic interference were mainly focused on analysing the
photoelectron energy spectrum. We show our results of  photoelectron energy spectra in
Fig.~\ref{fig1} at four different laser intensities. The results from the dipole approximation are consistent with previous studies,
i.e., the dynamic interference does appear for pulse intensity
above $10^{18}~\mathrm{W/{cm}^2}$ even for the ground state of hydrogen,
due to the onset of the atomic stabilization~\cite{Jiang2018}.
Atomic stabilization can be involved if the laser intensity is larger than $c{\omega}^4/8\pi$ for the atomic hydrogen~\cite{Jiang2018}.
In addition, we notice that the results from the nondipole calculation can perfectly agree with
those from the dipole calculations for laser intensity below $  10^{18}$ W/cm$^2$, see Figs.~\ref{fig1}(a)~and~\ref{fig1}(b).
However, obvious differences can be observed for higher laser intensities, see Figs.~\ref{fig1}(c)~and~\ref{fig1}(d).
For present photon energy ($\sim50$~eV), when the laser intensity is in the order of $10^{19}$ W/cm$^2$, one is approaching the limit of the dipole approximation (see, e.g. Fig.4 in Ref.~\cite{Reiss2014b}) and the nondipole effects may be non-negligible.
One can see from these results that the inclusion of the nondipole corrections can significantly
suppress the peak  splitting in the energy spectrum. Nevertheless, one can not draw the conclusion that the nondipole effects reduce the dynamic interference, as we will show below.

\subsection{Momentum shifts  in the rings of the dynamic interference}

\begin{figure}
  \centering
 \includegraphics[width=0.9\linewidth]{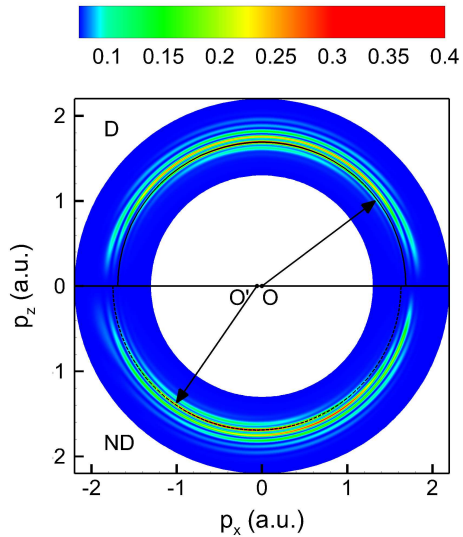}\\
  \caption{(Color online) The photoelectron momentum distribution $D(\mathbf{p})$ in the  $p_x$-$p_z$ plane ($p_y = 0$) plane at $I_0 = 1.0 \times 10^{19}~\mathrm{W/{cm}^2}$. The upper~(lower) half plane is the dipole~(nondipole) result with a black solid~(dashed) semi-circle as a reference to highlight the momentum shift. } \label{fig2}
\end{figure}

In order to check whether the dynamic interference is destroyed by the nondiple effects, we turn to examine the differential distributions of photoelectrons in the momentum  plane of $p_x$-$p_z$.  In Fig.~\ref{fig2}, we compare such momentum distributions from the dipole~(upper half-plane) and nondipole~(lower half-plane) calculations at the intensity of $1\times 10^{19}~\mathrm{W/{cm}^2}$.

Quite different from the gradual disappearance of the  interference oscillations in the energy spectra of Figs.~\ref{fig1}(c)~and~\ref{fig1}(d),
 one can see from Fig.~\ref{fig2} that the dynamic interference patterns are still present in the momentum space under the same laser parameters.
As mentioned earlier, the inclusion of the nondipole terms in the Hamiltonian will break the cylindrical symmetry in the photoelectron distributions. Indeed,
a careful examination for the lower half-plane can distinguish an asymmetric angular distribution in the
positive and negative direction of the laser propagation, i.e., the spectrum is no more symmetric about the line of the laser polarization $z$-axis.
The ring structures in the upper half plane of Fig.~\ref{fig2} precisely concenter at zero~(marked as the origin point O), while the centers of those ring structures in the nondipole result in the lower half-plane of Fig.~\ref{fig2} are roughly shifted towards the  opposite direction of the laser propagation~(marked as the origin point O$^\prime$).

Such momentum shifts explain why the peak splitting in the angle-integrated energy spectrum is gradually smoothed out.
 From the TDSE calculations, we can extract the value of the center move from the point O to O$^\prime$ as the momentum shift $\Delta p$ at different laser intensities, which are  shown by red solid circles in Fig.~\ref{fig3}(a).
One can see that the higher the laser intensity is, the larger the momentum shift is. The negative sign of $\Delta p$ means that the interference rings in the momentum space shift in the opposite direction  of the laser propagation.

One expects, without considering the momentum shifts, the dynamic interference would still be present in the energy spectrum. To show this, in Fig.~\ref{fig3}(b), we {\it artificially} move the shifted center O$^\prime$ of the nondipole interference patterns back to zero and recalculate
the angle-integrated energy spectrum $D(E)$.  Indeed, we find that the disappeared interference oscillations in Fig.~\ref{fig1}(d) can be largely retrieved and
the adjusted nondipole results  agree with the dipole results rather well.  From this respect,  we can deduce that the dynamic interference process is not prohibited by the magnetic force  for  present laser parameters and the main nondipole effects should be the momentum shift of the interference patterns, whose underlying mechanism will be discussed in the following subsection.

\subsection{Analytic expressions for the momentum shift}

 To see more clearly the nondipole effects, it's better to examine the momentum distributions in the polar coordinates of the $p_x$-$p_z$ plane~($p_y=0$).  As is shown in Fig.~\ref{fig4a}(a), the dipole  results are identical for the electrons with $p_x>0$ and $p_x<0$. However,  the angular distributions are drastically different for electrons with $p_x>0$ and those with $p_x<0$, as is respectively shown in Figs.~\ref{fig4a}(b)~and~\ref{fig4a}(c): the momentum is shifted toward a smaller or a larger value.  In the following, we will derive an analytical expression for the momentum shift.

\begin{figure}
  \centering
 \includegraphics[width=0.9\linewidth]{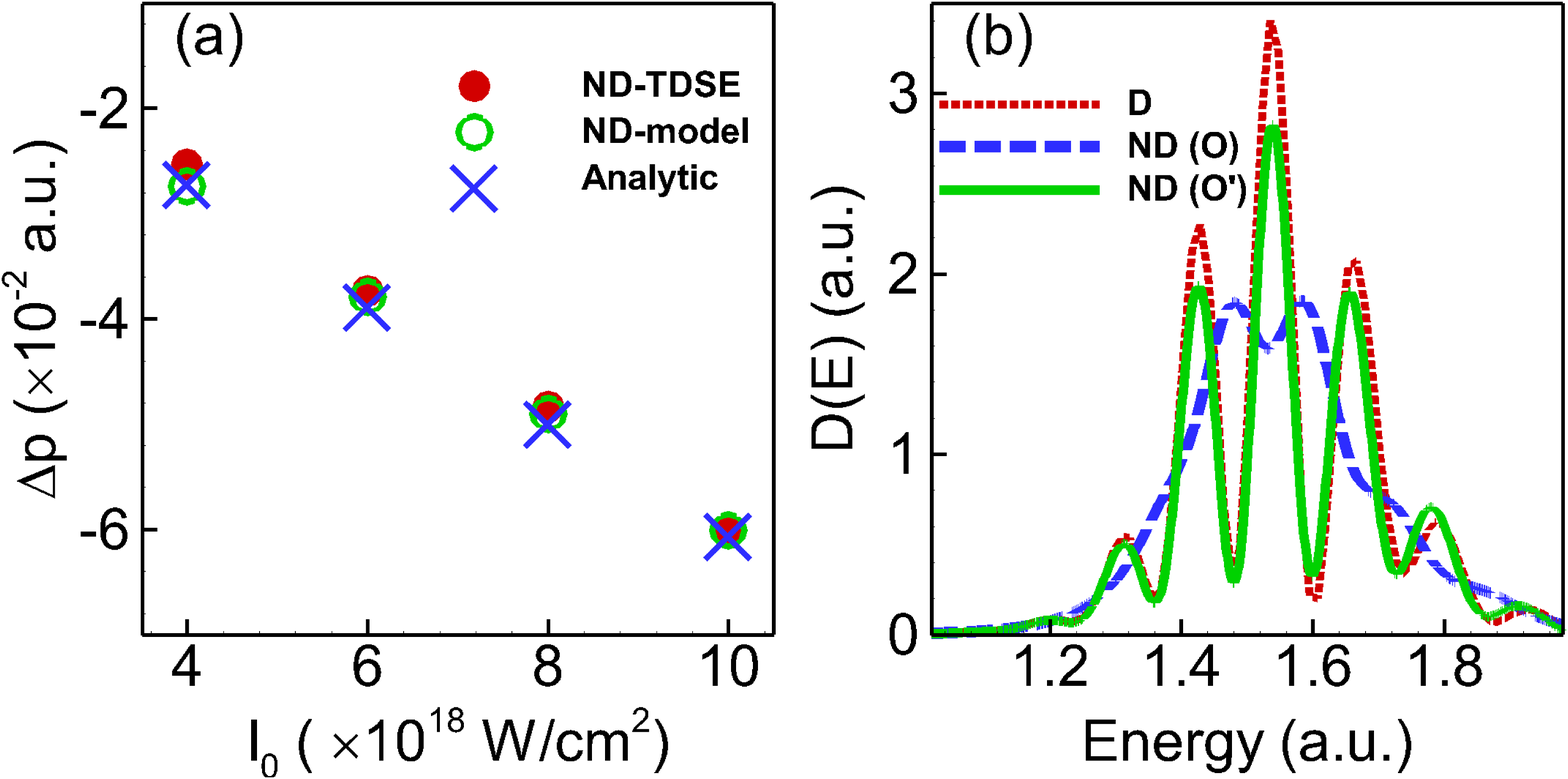}\\
  \caption{(Color online) (a) The momentum shift $\Delta p$ at various laser intensities: results in red solid circles (green open circles) are values extracted from the momentum distributions calculated by the nondipole TDSE (the semi-analytical model);  the blue crosses refer to the analytic results from Eq.~(\ref{deltap}). In (b), angle-integrated energy spectra $D(E)$ are shown: for the dipole result as a red dotted line,  and for the nondipole  result respectively calculated before/after moving the center O$^\prime$ back to zero by a blue dashed/green solid line.} \label{fig3}
\end{figure}

The dynamic interference origins from the AC-Stark shift of the initial ground state and
the final continuum state. Models based on the dipole approximation can
nicely explain the multi-peak structures in the  energy spectrum~\cite{Demekhin2012, Jiang2018}.
In the following, we are able to provide a physical explanation for the momentum shifts by including
the nondipole corrections into our previous semi-analytical model~\cite{Jiang2018} and presenting an analytic expression for the momentum shift.

\begin{figure}
  \centering
 \includegraphics[width=0.9\linewidth]{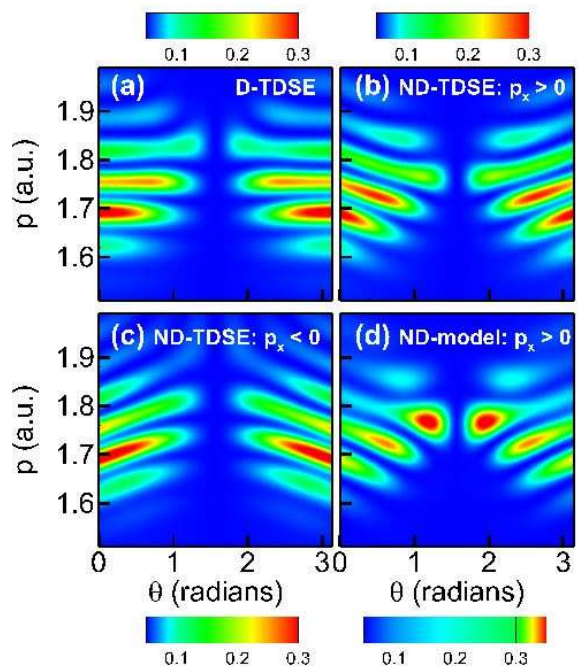}\\
  \caption{(Color online) The 2D photoelectron momentum distributions $D(\mathbf{p})$ in   the $p_x$-$p_z$ plane  ($p_y = 0$) at $I_0 = 1.0 \times 10^{19}~\mathrm{W/{cm}^2}$ are shown for: (a) the dipole TDSE calculation, (b) the nondipole TDSE calculation for $p_x > 0$, (c) the nondipole TDSE calculation for $p_x < 0$, and (d) semi-analytical model results for $p_x > 0$.} \label{fig4a}
\end{figure}

It is too complicated  to obtain an accurate expression for the AC-stark shift of
the continuum state, but it is not difficult to get an analytic
expression by neglecting the Coulomb potential in the Hamiltonian.
The solution of the TDSE without considering the Coulomb potential
is referred to the Volkov state.  A Volkov state is a reasonable
approximation for  the continuum state when the laser intensity is high enough, as those considered in the present work.
In addition, since the nondipole terms are only  a small correction to the dipole interacting Hamiltonian, one expects that the magnitude of the transition matrix from the ground to the continuum state will not change much. On the contrary, for the dynamic interference discussed in this work, the phase of the transition matrix will play a much more important role.  Therefore, let us satisfy ourselves by only including the nondipole modification in the phase.

The  Volkov phase in the velocity gauge can be expressed as~\cite{Joachain2011,He2017}
\begin{equation}
\Phi(t)=\phi_0(t)+\phi_1(t),
\end{equation}
where
\begin{equation} \label{dp}
\phi_0(t)=\int_{-\infty}^t \left[ \frac{p^2}{2}+p_zA(\tau) \right] d\tau
\end{equation}
is the dipole Volkov phase, and
\begin{equation} \label{np}
\phi_1(t)=\int_{-\infty}^t \frac{p_x}{c}\left[p_zA(\tau)+\frac{A^2(\tau)}{2}\right]d\tau
\end{equation}
is the nondipole correction for the Volkov phase. In Eq.~(\ref{np}), we
have neglected the small terms  $\propto\frac{1}{c^2}$ and the space-dependent phases~\cite{He2017}, which
have negligible influences on the interference.

By including the nondipole phase term into the semi-analytical model previously developed for  the dynamic interference, the momentum shift observed in the TDSE calculations is nicely reproduced, as is shown in Fig.~\ref{fig4a}(d) for $p_x > 0$. From the satisfactory agreement between Figs.~\ref{fig4a}(b)~and~\ref{fig4a}(d), one can conclude that  the momentum shift does mainly come from the nondipole correction of the Volkov phase.

With the help of Eq.~(\ref{np}), we can obtain the nondipole correction for the instantaneous
AC-Stark energy shift in the continuum state to be
\begin{equation} \label{ep}
\delta E(t)=\frac{d\phi_1(t)}{dt}= \frac{p_x}{c}\left[p_zA(t)+\frac{A^2(t)}{2}\right].
\end{equation}
Considering the high-frequency oscillations in
the AC-Stark energy shift and the laser envelope varying slowly compared to the
laser cycle, one may need to average the response of the system to the
laser field over the laser cycle to arrive at a formulation
solely expressed in terms of the laser envelope~\cite{Baghery2017a, Jiang2018}.
Therefore a mean filter can be applied to obtain the smoothed AC-Stark energy shift $\overline{\delta E}(t)$ without high-frequency oscillations.
The dipole term from Eq.~(\ref{dp}) and the first term in the nondipole correction in Eq.~(\ref{ep}) do not contribute to $\overline{\delta E}(t)$
because the time average of $A(t)$ is zero.
$\overline{\delta E}(t)$ only comes from the second term in the right hand of Eq.~({\ref{ep}})
and this   nondipole energy shift will affect the photoelectron momentum according to the energy conversation equation,
\begin{equation}\label{pp}
\omega=\frac{p^2}{2}+\overline{\delta E}(t)+I_p-\overline{\delta E_0}(t),
\end{equation}
where  $\overline{\delta E_0}(t) $ is the
energy shift the ground state after the mean filter.

Considering the  atomic stabilization effects, the time point of the  maximal ejection rate is not    at the peak of the pulse, but at the time
when the instantaneous intensity is in the order of $10^{17}$ W/cm$^2$ for the present case.
So the corresponding instantaneous AC-Stark energy shift $\overline{\delta E}(t)$ is rather small compared to the photoelectron kinetic energy.
We believe the instantaneous energy shift $\overline{\delta E}(t)$ barely contributes to the momentum shift mentioned above.

\begin{figure}
  \centering
 \includegraphics[width=0.9\linewidth]{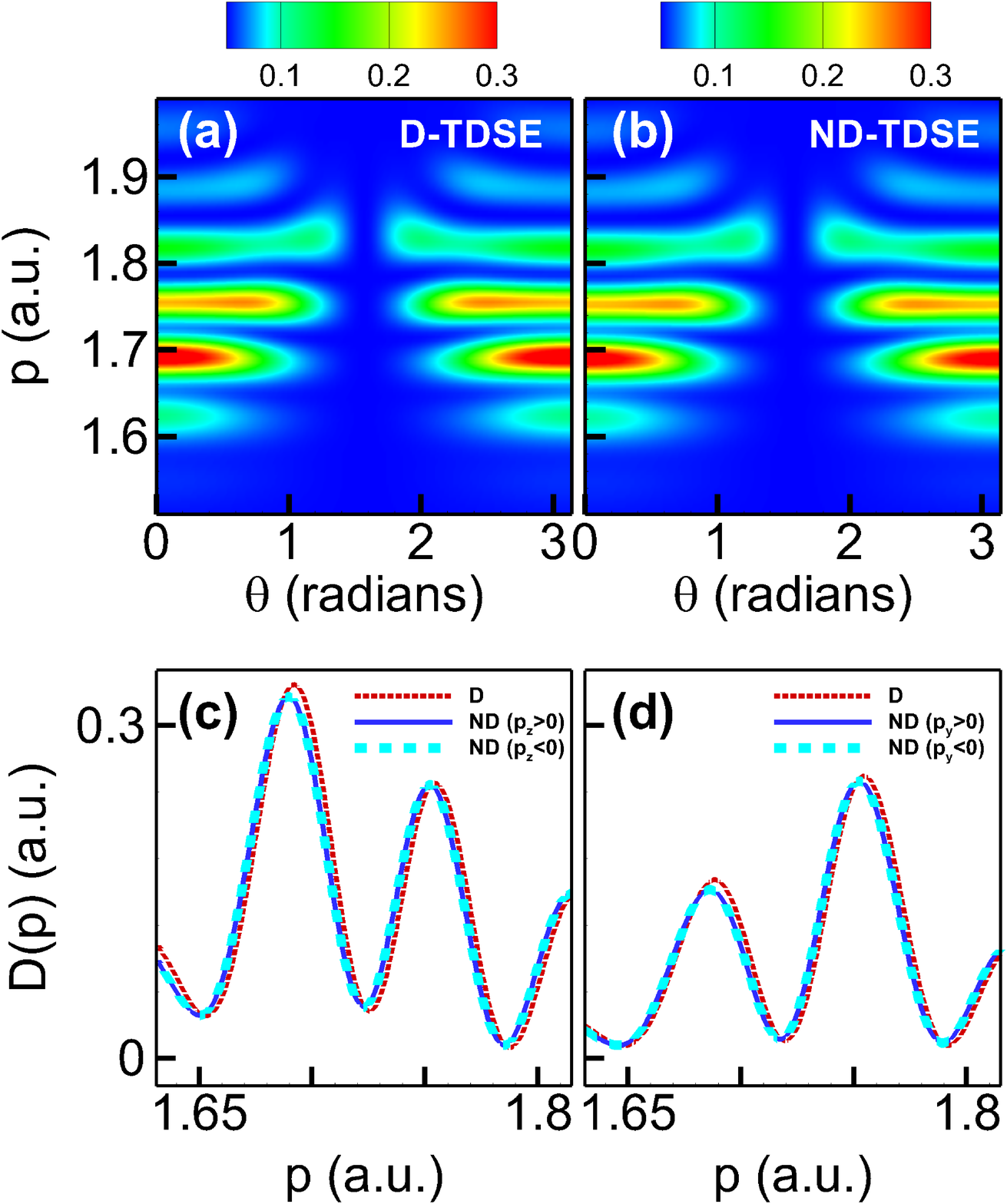}\\
  \caption{(Color online) The upper row shows the 2D photoelectron momentum distribution $D(\mathbf{p})$ in the $p_y$-$p_z$ plane ($p_x = 0$) at $I_0 = 1.0 \times 10^{19}~\mathrm{W/{cm}^2}$: (a) the dipole TDSE calculation for $p_y > 0$, and (b) the nondipole TDSE calculation for $p_y >0$. The lower row shows both the dipole and the nondipole $D(p)$ result for photoelectrons with $p_x = 0$: (c) along the laser polarization direction,  and (d) along the direction of $\theta = \pi/4$. } \label{fig5}
\end{figure}

Then we turn to the phase difference between two electron wave packets
ejected in the rising edge and the falling edge, which will interfere with each other and result in the fringe patterns. The ejection time points $t_1$ (in the rising edge) and $t_2$ (in the falling edge) of the photoelectron with momentum $\bf{p}$ can be obtained by solving Eq.~({\ref{pp}}).
In the period between the time point $t_1$ and $t_2$, the electron emitted at $t_1$ jumps to the Volkov state while the electron emitted at $t_2$ remains in the ground state.
Due to the nondipole correction to the Volkov phase, a nondipole term needs to be added only
to the phase of the electron emitted at $t_1$ in this period, which is equal to ${\phi}_1(t_2)-{\phi}_1(t_1)$, see Eq.~(\ref{np}).
Therefore the nondipole correction to the phase difference between the two electrons ejected at $t_1$ and $t_2$ emerges during this period and equals  ${\phi}_1(t_1)-{\phi}_1(t_2)$.
As a result of the high-frequency oscillation, the phase difference of the two electrons induced by the first term in Eq.~(\ref{np}) is negligible.
Therefore the nondipole term added to the phase difference can be expressed as,
\begin{equation} \label{phasediff}
\Delta \phi=-\frac{A_0^2p_x}{2c}\int_{t_1}^{t_2} g^2(t)dt,
\end{equation}
in which the time integration of $A^2(t)$ is given by $A_0^2 \int_{t_1}^{t_2} g^2(t)dt $, where $A_0$ is the peak value of the vector potential of the whole pulse.

Due to the positive AC-Stark shift of the ground state, the phase of the secondly ejected electron is larger than that of the firstly ejected electron when the two electrons interfere.
The larger the ejection time difference is, the greater the phase difference will be.
For positive (negative) $p_x$, the nondipole correction will decrease (increase) the phase difference, moving the dipole interference fringe to the direction of smaller (larger) momentum value.
This finally leads to a momentum shift which is opposite to the laser propagation direction.
 If the change of the phase difference equals 2$\pi$, the peak positions will be shifted by the fringe spacing $\Delta p_0$. For $\Delta\phi$ much smaller than 2$\pi$, we assume the momentum shift depends on $\Delta\phi$ linearly. Then the shift of the peak positions in the momentum spectra can be estimated by,
\begin{eqnarray} \label{peak}
\nonumber \Delta p_\text{peak}&&= \frac{\Delta \phi}{2\pi}\Delta p_0            \\
&&=- \frac{2 \int_{t_1}^{t_2} g^2(t)dt}{{\omega}^2 c^2 } I_0 \Delta p_0 p_x.
\end{eqnarray}
where $I_0$ is the laser peak intensity, $\Delta p_0$ is the fringe spacing,  and $p_x$ is the momentum component along the laser propagation direction.

In Eq.~(\ref{peak}), it can be seen that the nondipole peak shift
is zero if $p_x=0$. This is consistent with our TDSE results  in Figs.~\ref{fig5}(a)~and~\ref{fig5}(b), where the momentum distributions in the $p_y$-$p_z$ plane are shown for  $p_x=0$. In this case, the nondipole effects can indeed be neglected, as can be observed more clearly by looking a line cut in Figs.~\ref{fig5}(c)~and~\ref{fig5}(d). However, for $p_x\ne 0$, the nondipole effects will become significant and the momentum shift will depend on the sign of $p_x$.   According to Eq.~(\ref{peak}), $\Delta p_{\text{peak}}$ is negative for electrons with $p_x>0$ and positive
for $p_x<0$. This can explain the different trends of shifts in the momentum space for electrons with
positive and negative $p_x$ in Figs.~\ref{fig4a}~(b)~and~\ref{fig4a}(c).

In Fig.~\ref{fig2}, the peak positions in the nondipole calculation are  roughly located at  circles
whose center has been shifted away from the origin of coordinates.
This shift of center can be approximately deduced from Eq.~(\ref{peak}) to be,
\begin{equation} \label{deltap}
\Delta p \approx -\frac{2 \int_{t_1}^{t_2} g^2(t)dt}{{\omega}^2 c^2 } p \Delta p_0 I_0.
\end{equation}
For present laser pulses with high intensities, the time points $t_1$ and $t_2$  of the dominated ejection are located close to the
tails of the laser pulse due to the atomic stabilization.
Hence in the estimation of Eq.~(\ref{deltap}), it is possible to approximate $\int_{t_1}^{t_2} g^2(t)dt$  by $\int_{-\infty}^{\infty} g^2(t)dt$.
The predictions from Eq.~(\ref{deltap})  are presented as blue crosses in Fig. \ref{fig3}(a), and one
can see that they
agree very well  with the shifts respectively extracted from the TDSE  and the semi-analytical model.

\section{Conclusions}

\par In summary, by solving the 3D time-dependent Schr\"odinger equation within/beyond the dipole approximation, we have observed clear dynamic interference exhibited in the ionization of the hydrogen ground state by a super-intense high-frequency laser pulse. The oscillation patterns in the photoelectron energy spectra are controlled by the laser intensity and gradually wiped out in the nondipole calculations.  However, through an all-round analysis of the differential momentum distributions of photoelectrons, we find that  the smoothing of the peak splitting in the  energy spectra is due to momentum shifts in the opposite direction of the laser propagation, which are proportional to the  peak laser intensity. By developing a semi-analytical model including the nondipole Volkov phase, all the features observed in the TDSE calculations can be well reproduced. Through an analytic expression for the momentum shift, we point out that the physical origin of this momentum shift can be mainly attributed to the nondipole phase difference between the two electron wave packets ejected in the rising edge and the falling edge. With the fast development of the strong short-wavelength lasers, the predicted phenomena may be observed in the future.

\vspace*{0.1cm}
\begin{acknowledgments}
 This work is supported by the National Natural Science Foundation of China~(NSFC) under Grant Nos. 11725416, 11574010 and 11747013, and by the National Key R\&D Program of China~(Grant No. 2018YFA0306302).  L.Y.P. acknowledges the support by the National Science Fund for Distinguished Young Scholars.
\end{acknowledgments}

\bibliography{nddynamic}

\end{document}